\documentclass[apjl]{emulateapj}
\usepackage{graphicx}
\usepackage{color}
\usepackage{amsmath}
\usepackage{xspace}

\usepackage[breaklinks,colorlinks, urlcolor=blue,citecolor=blue,linkcolor=blue]{hyperref}

\shortauthors{Barclay, et al. 2014}
\shorttitle{Importance sampling and Kepler-296}

\begin{document}

%
\def\ltsima{$\; \buildrel < \over \sim \;$}
\def\lsim{\lower.5ex\hbox{\ltsima}}
\def\gtsima{$\; \buildrel > \over \sim \;$}
\def\gsim{\lower.5ex\hbox{\gtsima}}
\def\K{\emph{Kepler}}

\newcommand{\numax}{\mbox{$\nu_{\rm max}$}}
\newcommand{\Dnu}{\mbox{$\Delta \nu$}}
\newcommand{\dnu}[1]{\mbox{$\delta \nu_{#1}$}}
\newcommand{\muHz}{\mbox{$\mu$Hz}}
\newcommand{\kep}{\mbox{\textit{Kepler}}}

\newcommand{\msun}{\mbox{$M_{\sun}$}}
\newcommand{\rsun}{\mbox{$R_{\sun}$}}
\newcommand{\lsun}{\mbox{$L_{\sun}$}}
\newcommand{\rearth}{\mbox{$R_{\oplus}$}}

\newcommand{\rp}{\mbox{$1.68^{+0.28}_{-0.28}$}}
\newcommand{\teq}{\mbox{$283\pm17$}}

\newcommand{\rpone}{\mbox{$2.67^{+0.45}_{-0.44}$}}

\newcommand{\starname}{Kepler-XX\xspace}
\newcommand{\planetb}{Kepler-XXb\xspace}
\newcommand{\planetc}{Kepler-XXc\xspace}
\newcommand{\planetd}{Kepler-XXd\xspace}
\newcommand{\planete}{Kepler-XXe\xspace}
\newcommand{\planetf}{Kepler-XXf\xspace}

\newcommand{\logg}{\mbox{$\log{g}$}}
\newcommand{\feh}{\mbox{$\rm{[Fe/H]}$}}
\newcommand{\id}{\mbox{KIC\,8311864}}

\newcommand{\teff}{\mbox{$T_{\textrm{eff}}$}}

\newcommand{\be}{\begin{equation}}
\newcommand{\ee}{\end{equation}}

\def\lta{\,\raise 0.3 ex\hbox{$ < $}\kern -0.75 em
 \lower 0.7 ex\hbox{$\sim$}\,}
\def\gta{\,\raise 0.3 ex\hbox{$ > $}\kern -0.75 em
 \lower 0.7 ex\hbox{$\sim$}\,}

%

\title{THE FIVE PLANETS IN THE KEPLER-296 BINARY SYSTEM ALL ORBIT THE PRIMARY: 
A STATISTICAL AND ANALYTICAL ANALYSIS}

\author{
Thomas Barclay$^{1,2}$,
Elisa V. Quintana$^{1,11}$,
Fred C. Adams$^{3}$,
David R. Ciardi$^{4}$,
Daniel Huber$^{5,6,7}$,
Daniel Foreman-Mackey$^{8}$,
Benjamin T. Montet$^{9, 10}$,
Douglas Caldwell$^{1,6}$,
}

\newcommand{\ron}{}
\newcommand{\new}{}

\slugcomment{Accepted version \today, by ApJ}

\altaffiltext{1}{NASA Ames Research Center, M/S 244-30, Moffett Field, CA 94035, USA}
\altaffiltext{2}{Bay Area Environmental Research Institute, 625 2nd St. Ste 209 Petaluma, CA 94952, USA}
\altaffiltext{3}{Department of Physics, University of Michigan, 450 Church Street, Ann Arbor, MI 48109, USA}
\altaffiltext{4}{NASA Exoplanet Science Institute, California Institute of Technology, MC 100-22, Pasadena, CA 91125, USA}
\altaffiltext{5}{Sydney Institute for Astronomy (SIfA), School of Physics, University of Sydney, NSW 2006, Australia}
\altaffiltext{6}{SETI Institute, 189 Bernardo Avenue, Mountain View, CA 94043, USA}
\altaffiltext{7}{Stellar Astrophysics Centre, Department of Physics and Astronomy, 
Aarhus University, Ny Munkegade 120, DK-8000 Aarhus C, Denmark}
\altaffiltext{8}{Center for Cosmology and Particle Physics, Department of Physics, New York University, 4 Washington Place, New York, NY 10003, USA}
\altaffiltext{9}{Cahill Center for Astronomy and Astrophysics, California
Institute of Technology, 1200 E. California Blvd.,
Pasadena, CA 91125, US}
\altaffiltext{10}{Harvard-Smithsonian Center for Astrophysics, 60 Garden
Street, Cambridge, MA 02138, USA}
\altaffiltext{11}{NASA Senior Fellow}


\begin{abstract}

Kepler-296 is a binary star system with two M-dwarf components separated by 0.2''. Five transiting planets have been confirmed to be associated with the Kepler-296 system; given the evidence to date, however, the planets could in principle orbit either star. This ambiguity has made it difficult to constrain both the orbital and physical properties of the planets. Using both statistical and analytical arguments, this paper shows that all five planets are highly likely to orbit the primary star in this system. We performed a Markov-Chain Monte Carlo simulation using a five transiting planet model, leaving the stellar density and dilution with uniform priors. Using importance sampling, we compared the model probabilities under the priors of the planets orbiting either the brighter or the fainter component of the binary. A model where the planets orbit the brighter component, Kepler-296A, is strongly preferred by the data. Combined with our assertion that all five planets orbit the same star, the two outer planets in the system, Kepler-296 Ae and Kepler-296 Af, have radii of $1.53\pm0.26$ and $1.80\pm0.31$ $R_\oplus$, respectively, and receive incident stellar fluxes of $1.40\pm0.23$ and $0.62\pm0.10$ times the incident flux the Earth receives from the Sun. This level of irradiation places both planets within or close to the circumstellar habitable zone of their parent star.

\end{abstract}


\section{Introduction}

More than half of the stars in our Galaxy are components of multiple star systems \citep{duquennoy91,raghavan10}. From the many hundreds of exoplanets already discovered \citep[e.g.,][]{rowe14}, it has been estimated that as many as 40 -- 50\% may orbit a star with a bound stellar companion \citep{horch14}. In addition, provided that the stellar components are well separated ($a>10$ AU), there appears to be no suppression of the planet occurrence rates for binary star systems \citep{wang14}.

For about half of the transiting exoplanet host stars that are members of a binary star system, however, establishing which stellar component the planet orbits is not trivial \citep{horch14}. Often binary stars cannot be resolved as separate stars.  Radial velocity observations of a planet in a binary star system can help identify which of the stars is the exoplanet host (except perhaps for an equal mass binary), and in some cases even reveal the presence of a stellar companion \citep{gilliland13}. Radial velocity observations are not always available, however, thus it is often the case that we know that a transiting planet exists but that the planet's host star remains uncertain.

The Kepler-296 system, which harbors five small planets, is a prime example of exoplanets in a binary system. The system consists of two stars separated by 0.22'' \citep{horch12} with a brightness difference of 1.72 mag at 692 nm. \citet{lissauer14} and \citet{star14} have reported that these two stars are highly likely to be bound M-dwarfs. The Kepler pipeline detected five transiting planet signals \citep{batalha12,tenenbaum13,tenenbaum14,burke14} which were designated Kepler Object of Interest numbers KOI-1422.01 to .05\footnote{The KOI numbers for this system do not increase monotonically with orbital period but represent the order in which they were detected. Hence, KOI-1422.05 has a shorter orbital period than KOI-1422.04.} by the Kepler team. These five candidates were later verified as planets through multiplicity arguments \citep{rowe14,lissauer14}\footnote{We note that both \citeauthor{rowe14} and \citeauthor{lissauer14} report a planet with an orbital period of 3.62 d. However, we believe that this planet has an orbital period precisely three times longer at 10.86 d. We discuss this discrepancy further in Section~\ref{sec:nomen}.}. Only the planetary nature of these candidates was verified, however, and the assignment of each planet to a host star was not determined, nor whether all of the planets orbited the same star. In other words, \citet{lissauer14} found valid solutions for any of the planets to orbit any of the stars since the planets remained small ($<$4 \rearth) in all cases. More recently, \citep{torres15} performed a rigorous statistical analysis of false positive probabilities to independently validate the Kepler-296 system, showing through transit modeling that the planets likely orbit the same star and subsequently established planet and orbital parameters for two of the planets that were found to orbit in the habitable zone (which was the primary focus of their study).

Herein we take a novel approach to determining the host star of the five transiting planets. We look what the transit light curves tell us about the planet's host star and then assess which star the planets' are more likely to orbit based our understanding of the physical properties of the two stars in the binary. We then use these analyses to constrain the physical parameters of these planets


\section{Ground-based observations of Kepler-296}
\label{sec:ground}
\begin{figure}
\includegraphics[width=0.47\textwidth]{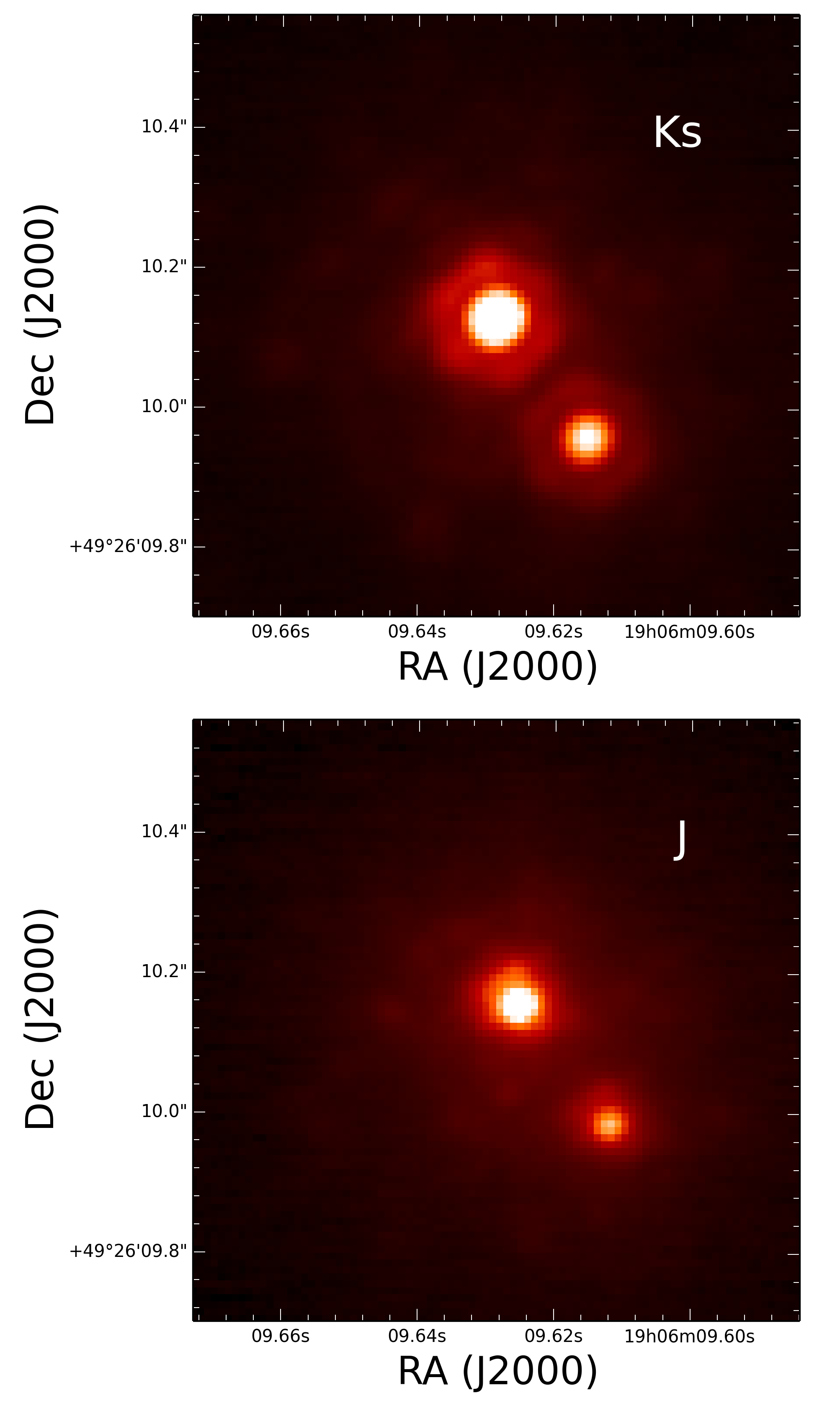}
\caption{NIRC-2 images of Kepler-296 in $Ks$ and $J$ filters show two stars separated by 0.217''. The two images are scaled appropriately to account for differences in exposure time. The magnitude difference between the two stars is $\Delta J=1.10$ and $\Delta Ks=1.14$.}
\label{fig:ao}
\end{figure}

We observed Kepler-296 using the NIRC-2 instrument on the Keck-II telescope using the Laser Guide Star Adaptive Optics (LGS-AO) system on 2013-08-08. We obtained a total integration time of 80 s with the $Ks$ filter and 36 s in $J$ band. As shown in Figure~\ref{fig:ao}, two stellar components are clearly resolved in the AO images. The flux ratio between the two stars in the $J$-band image is $1.10\pm0.04$ and in the $Ks$-band image is $1.14\pm0.04$ (brightnesses are shown in Table~\ref{tab:stellar}).

\begin{table}
\centering
\caption{Summary of properties derived from AO data}\label{tab:stellar}
\begin{tabular}{l l l}
Property & Primary & Secondary \\
\hline
J				&	$13.73\pm0.03$	&	$14.83\pm0.03$\\
Ks				& 	$12.93\pm0.03$	&	$14.07\pm0.03$\\
(J-Ks)			&	$0.80\pm0.04$		&	$0.76\pm0.04$\\
$\Delta$R.A. (arcsec)&					&	-0.130\\
$\Delta$Dec. (arcsec)&					&	-0.174\\
$\Delta\theta$ (arcsec)&					&	0.217\\
\hline
\end{tabular}
\end{table}

Kepler-296 was observed as part of a campaign to obtain infrared spectra of cool Kepler Objects of Interest (KOIs) \citep{muirhead12,muirhead14}. They reported a stellar effective temperature, $T_\textrm{eff} = 3520\pm70$ K and an iron abundance [Fe/H] = $-0.08\pm0.14$ dex. 

We reanalyzed the spectrum from \citet{muirhead14} using the spectral fitting technique created by \citet{covey10} and improved by \citet{rojas-ayala12} but with a two spectrum model fit applied to account for the two stars in Kepler-296 system. This is a similar strategy to that employed and described by \citet{montet14}. Specifically, we extract information on the temperature and metallicity of each star from observations of the $K_s$ band sodium doublet, calcium triplet, and absorption due to water opacity (the ``$H_2O-K2$ index''). In the fitting routine we fixed the $\Delta$J-band magnitude difference between the two stars to the values we measured using the NIRC-2 AO data. This analysis yielded temperatures for the two stars of $3435 \pm 58$ K and $3770 \pm 128$ K and a surface gravity of $4.89 \pm 0.04$ dex and $4.74 \pm 0.04$ dex, for the secondary and primary star, respectively. 

When applying this method to combined-light spectra of two stars, there is a degeneracy between the allowed temperatures and metallicities of each star. Such degeneracy is reduced if we assume both stars have the same metallicity, as would be expected for a close binary pair. Assuming the stars are co-eval from the same cloud, we then fit for one metallicty, measuring [M/H] = $-0.05 \pm 0.29$. We note that the parameters found here are somewhat cooler than \citet{star14} who used HST photometry to determine their stellar properties. We have opted to use our values which are derived spectroscopically to the photometric properties previously reported.

To estimate radii, densities and the flux ratio of the two stars we used
a grid of Dartmouth isochrones \citep{dotter08}. The grid is based on
solar-scaled alpha-element abundances, and was interpolated to a stepsize of 0.01\,\msun\ 
in mass and 0.02\,dex metallicity using the interpolation tool provided in the Dartmouth 
database\footnote{The Dartmouth isochrone interpolation tool is available from \url{http://stellar.dartmouth.edu/models/programs/iso_interp_feh.f}.}.

We used \teff, \logg\ and \feh\ of the primary
to infer stellar properties from the isochrone grid using a Markov-Chain 
Monte-Carlo algorithm. We adopted
uniform priors in mass, age, and metallicity. For ease of computation samples in age and 
metallicity were drawn in discrete steps corresponding to the sampling of the model grid 
(0.5\,Gyr in age and 0.02\,dex in \feh). For each sample of fixed age and metallicity, 
we interpolated the grid in mass to derive \teff\ and \logg, 
which are used to evaluate the likelihood function given the spectroscopic values for 
\teff, \logg\ and \feh.
In addition to \teff\ and \logg, we interpolate in mass to derive 
absolute magnitudes in the Ks ($M_{Ks}$) and Kepler ($M_{Kp}$) bandpasses for the primary. 
For the secondary, we added at each step the $Ks$ contrast from AO imaging 
with a Gaussian random error to $M_{Ks}$ derived for the primary, and 
interpolated the grid in $M_{Ks}$ to derive \teff, mass, 
\logg\ and $M_{Kp}$ for the secondary. Note that by
fixing the secondary parameters to the primary we assume that both 
stars have the same age, metallicity, and distance. 

\begin{figure}
\includegraphics[width=0.47\textwidth]{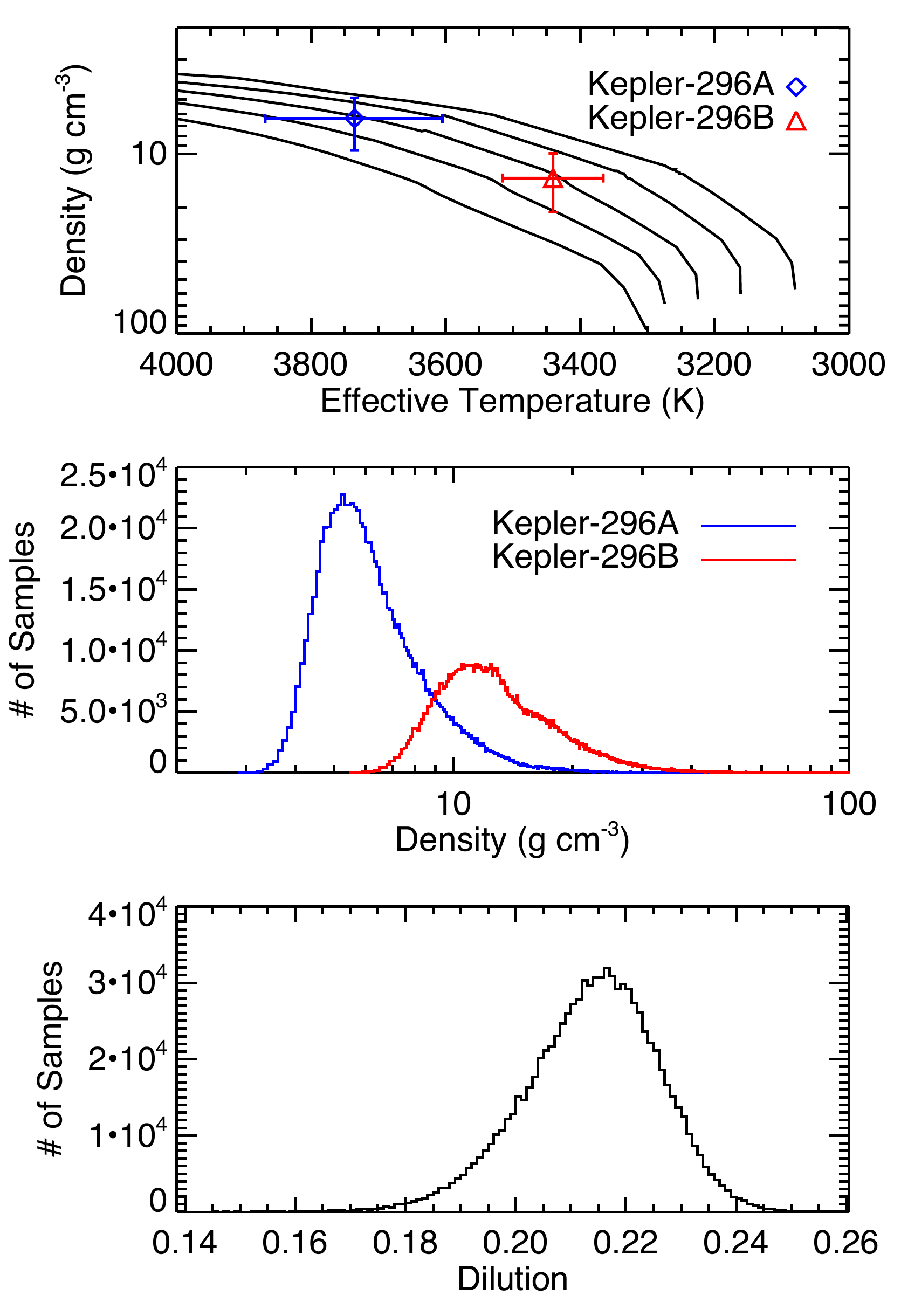}
\caption{Top panel: Dartmouth isochrones with an age of 6\,Gyr and metallicities of 
$=-0.64, -0.34, -0.06, 0.22$ and 0.48, roughly corresponding to the 2\,$\sigma$ 
error bar from spectroscopy. The positions of Kepler-296A and B based on the median of the 
MCMC posteriors are shown as red diamond and blue triangle, respectively. 
Middle panel: Posteriors for the density of both components. Bottom panel: 
Posterior of the dilution in the Kepler bandpass.}
\label{fig:hrd}
\end{figure}

We calculated $10^6$ iterations (discarding the first 10\% as burn-in), and verified 
that the results are unaffected by the choice of initial guesses. 
The resulting MCMC chains provide stellar properties and 
$Ks$ and Kepler bandpass contrasts for both components (Table~\ref{tab:stellar}). 
Figure \ref{fig:hrd} shows 
both components in a \teff-density diagram, as well as the stellar density 
posteriors and the derived dilution in the Kepler bandpass. We note that the 
properties for the secondary from the isochrone fit ($\teff=3450\pm75$\,K, 
$\logg=4.93^{+0.09}_{-0.06}$) are in good agreement with the independent estimates from the 
spectroscopic analysis ($\teff=3435\pm58$\,K, $\logg=4.89\pm0.04$).

Radii of interior models for cool stars are well known to show offsets from empirical 
observations such as long-baseline interferometry \citep{boyajian12b} or eclipsing 
binaries \citep{lopez07,irwin09}. While recent models including magnetic fields effects 
on convection show partial success to reproduce observations \citep{feiden13}, 
empirical calibrations are commonly employed to 
avoid model-dependent offsets when estimating radii for cool planet host 
stars \citep{mann13}. 
Using the most recent $\teff-R-\feh$ relation from \citet{mann15} we derive empirical radii of 
$0.51$\rsun\ for Kepler-296\,A and $0.35$\rsun for Kepler-296\,B, which are 
$\sim$6-9\% ($0.5-0.7\sigma$) larger than the radii inferred from the isochrone grid. 
Importantly the fractional offsets are similar for both 
components and hence the densities of Kepler-296\,A and B 
should not be strongly affected by differential model-dependent offsets.


Kepler has 4'' pixels, therefore the two components which are separated by 0.217'' are unresolved in the Kepler data and can be treated as a point source. From our MCMC modeling we estimate a brightness difference in the Kepler bandpass of $\Delta$Kp=$1.41\pm0.08$ which is equivalent to saying that $78.5\pm1.2$\% of the light is coming from the primary and $21.5\pm1.2$\% from the secondary star in the binary. This is in good agreement with the results of \citet{gilliland15} and \citet{star14} who report Hubble Space Telescope observations that measure a brightness ratio of the two components of 5-to-1. Hereafter we define dilution as the proportion of total light coming from the other star, so the dilution of the primary star is $0.215\pm0.012$.

\begin{table}
\centering
\caption{Derived stellar properties}\label{tab:stellarprop}
\begin{tabular}{l l l}
Property & Primary & Secondary \\
\hline
\teff\ (K)			&	$3740^{+130}_{-130}$	&	$3440^{+75}_{-75}$		\\
\logg\ (dex)			& 	$4.774^{+0.091}_{0.059}$	&	$4.933^{+0.087}_{-0.063}$	\\
\feh\ (dex)		&	$-0.08^{+0.28}_{-0.30}$	&	$-0.08^{+0.28}_{-0.30}$	\\
Radius (\rsun)		&	$0.480^{+0.066}_{-0.087}$&	$0.322^{+0.060}_{-0.068}$	\\
Mass (\msun)		&	$0.498^{+0.067}_{-0.087}$&	$0.326^{+0.070}_{-0.079}$\\
$\rho$ (g/cc) 			&	$6.4^{+3.2}_{-1.5}$		&	$14^{+8}_{-4}$		\\
$\Delta$Kp			&	0					&	 $1.409^{+0.085}_{-0.070}$		\\
Dilution				&	$0.215\pm0.012$		&	$0.785\pm0.012$		\\
\hline
\end{tabular}
\end{table}

Our analysis of the stellar properties relies on the two stars being physically bound. There are compelling arguments in previous work demonstrating that the two stars are highly likely to form a physically bound binary. \citet{star14} used model isochrones and performed a numerical analysis to show that the chances of observing two stars with these colors by chance is extremely unlikely. Similarly, \citet{torres15} look at the stellar density in this region of the sky where Kepler-296 is and used Galactic models to predict the chance alignment of an unbound star is very small compared with a bound companion scenario. In addition, if these stars were unbound they this would likely result in two sets of stellar lines caused by differing radial velocities. However, \citet{torres15} find no additional lines of another star appear in a Keck/HIRES spectrum implying that the two stars have very similar radial velocities, as would be expected if they were bound.

\section{Kepler light curve modeling}
\label{sec:lightcurve}

We modeled the long cadence Kepler data \citep{quintana10,twicken10a} using a light curve model described by \citet{rowe14}, \citet{barclay13b} and \citet{quintana14} that comprises of limb darkened transits \citep{mandel02} of five planet and allows the planets to have eccentric orbits. The parameters we use to describe the model are the mean density of the star, a linear and a quadratic limb darkening coefficient, a photometric zero-point, and for each planet, the mid-point of the first transit, the orbital period of the planet, the impact parameter at time of mid-transit, the planet-to-star radius ratio and two eccentricity vectors $e\sin{\omega}$ and $e\cos{\omega}$ where $e$ is the eccentricity and $\omega$ is the argument of periastron. We also include an additional white noise term that is added in quadrature with the uncertainty reported in the Kepler data products. Finally, we include the dilution from the other star as a model parameter.

We used the Q1-Q15 Kepler light curves. This target falls onto the failed Module 3 which resulted in no data for this source being taken in Quarters 8, 12 and 16. We used pre-search conditioned light curve data \citep{twicken10b} which minimizes the instrumental signals. To remove astrophysical variability such as star spots we used a running median filter but weighted the transits zero in this filtering so as to avoid overly distorting the transit profiles.

We used \texttt{emcee}, an implementation of an affine invariant Markov Chain Monte Carlo algorithm \citep{goodman10,foreman13} to efficiently explore the posterior probability of our transit model. We have made the assumption that all the planets orbit a single star but that star could either be Kepler-296A or Kepler-296B (for a discussion of the validity of this assumption see Section~\ref{sec:validity}). In the sampling we assumed uniform priors on the mean stellar density between $10^{-4}$ and 200 g/cc and a dilution, $f$ that is uniform between 0 and 1 where the total light from the system is unity and the light from the transited star is $1-f$.


We use uniform priors on the photometric zero point, the transit mid-point times, the orbital period of the planets, the impact parameters are positive uniform as is the planet-to-star radius ratio. Following \citet{kipping13} we use a prior on the eccentricity of the orbits of the planets that takes the form of a beta distribution
\begin{equation}
P_{\beta} = \frac{1}{B(a,b)} e^{a-1} (1-e)^{b-1}
\end{equation}
where $e$ is the orbital eccentricity of each planet and $B$ is the beta function. \citet{rowe14} found that for multiple transiting planet systems the best fitting values for parameters ($a,b$) are $a$=0.4497 and $b$=1.7938 which we use in this work. In addition, we do not allow eccentricities that would allow for crossing orbits of any of the planets by excluding solutions where the periastron (or apastron) values are greater than (or less than) the semimajor axis of adjacent planets. We do, however, include a $1/e$ term in our likelihood function because our choice to sample in $e\sin{\omega}$ and $e\cos{\omega}$ (rather than $e$ and $\omega$) introduces a bias toward high values of $e$ if not modeled correctly \citep{eastman13}.

Limb darkening is poorly constrained in the regime of cool stars \citep{claret12, csizmadia13}, so we assume a uniform distribution of the two limb darkening parameters but do enforce priors that keep the stellar brightness profile physical \citep{burke08}.

In each simulation we used 800 chains with each chain taking 50,000 steps in the posterior probability. However, we discarded the first 10,000 steps in each chain for burn-in which leaves 32M samples of the posterior probability. 

\section{Model comparison} 
\label{sec:compare}

In our MCMC sampling we assumed uniform priors on stellar density and dilution. However, we are not ignorant of these parameters, indeed Section 2 describes our efforts to constrain these parameters. The reason for using uniform priors is that sampling from a bi-model parameter space which have well separated modes in a single model is difficult for standard MCMC algorithms \citep{kruschke14}, while performing two MCMC simulations would necessitate computing the marginalized likelihood which is a hard problem \citep{loredo99}.

\begin{figure}
\includegraphics[width=0.47\textwidth]{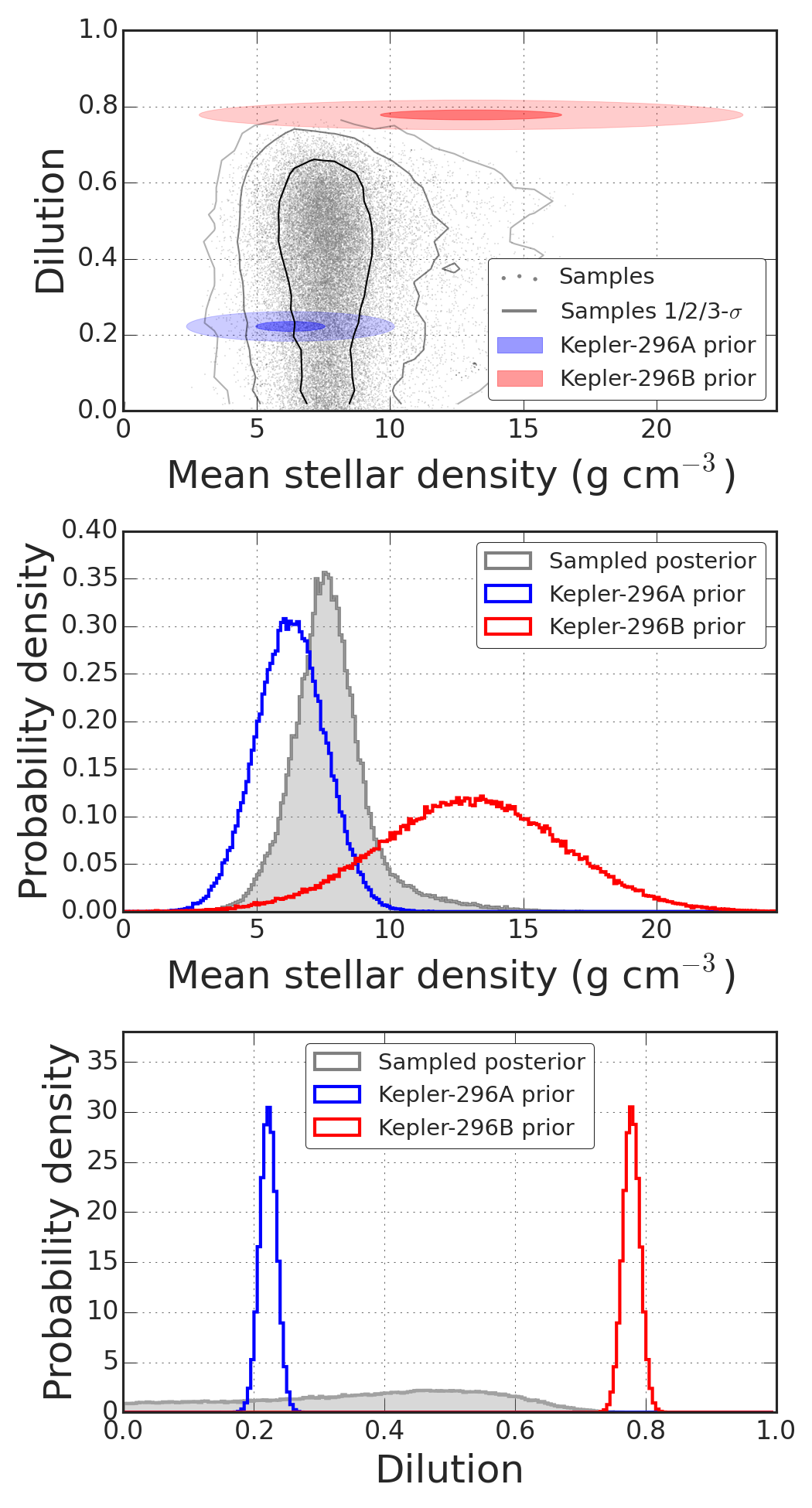}
\caption{The sampled posterior probability based on the Kepler light curve data and priors. The sampled posterior is shown in gray, the prior if the five planets orbit Kepler-296A is shown in blue and the prior if the planets orbit Kepler-296B in red. The upper panel shows the joint-distribution of dilution and mean stellar density. The sampled data is shown as gray dots and the 1, 2 and 3-$\sigma$ bounds of this sampling are shown as gray lines of decreasing opacity. The dark blue and red ellipses are the 1-sigma joint distribution of the prior for Kepler-296 A and B respectively. The fainter blue and red ellipses are the 3-$\sigma$ bounds. Note that the sampling of the dilution-$\rho$ distribution show negligible covariance. The central panel shows the mean stellar density marginal posterior distribution and the lower panel shows the dilution marginal posterior distribution. The Kepler-296A priors intuitively look more consistent with the observed Kepler light curve data than the Kepler-296B prior. We used importance sampling to quantify this intuition and we conclude with 99.9\% probability that the five planets orbit Kepler-296A.}
\label{fig:imp_sampling}
\end{figure}

We used a technique called importance sampling to re-sample the posterior under different priors than were used in the MCMC sampling. This is a well tested and used method when the posterior is difficult to sample directly \citep{hogg10}. Here we use importance sampling as a Bayesian evidence estimator. In this example we have two models which we label $\theta_1$ and $\theta_2$, with two different sets of priors that contain our prior knowledge of the dilution and density for the planets orbiting the brighter star and the fainter star, respectively. Let $\alpha$ be the model we actually sampled from which is uniform in density and dilution. So we have $K$ samples of $(\rho_k, f_k)$ where 
\begin{equation}
(\rho_k, f_k) \sim p(\rho,f | \bold{x},\alpha) = \frac{p(\rho,f|\alpha)p(\bold{x}|\rho,f)}{p(\bold{x}|\alpha)}
\end{equation}
$(\rho_k, f_k)$ is drawn from $p(\rho,f | \bold{x},\alpha)$ where $\bold{x}$ is the observed data. Now what we want to compute is the marginalized likelihood of the data under models $\theta_1$ and $\theta_2$, e.g.
\begin{eqnarray}
p(\bold{x}|\theta_n) &=& \int \mathrm{d}\rho\, \mathrm{d}f\, p(\bold{x},\rho,f|\theta_n) \\
&=& \int \mathrm{d}\rho\, \mathrm{d}f\, p(\rho,d|\theta_n)p(\bold{x}|\rho, f)
\end{eqnarray}
where $p(\rho,f|\theta_n)$ is the prior we want to enforce ($\theta_n$ can be either $\theta_1$ or $\theta_2$) and is based on the measured density $\rho_q$ and dilution $f_q$ with uncertainties $\delta \rho_q$, $\delta f_q$ we calculated in Section~\ref{sec:ground}, such that 
\begin{equation}
p(\rho, f|\theta_n) = N(\rho; \rho_q, \delta \rho_q)\, N(f; f_q, \delta f_q) 
\end{equation}
Using the posterior samples from our MCMC sampling, we can approximate the integral as
\begin{eqnarray}
p(\bold{x}|\theta_n)&=&\int \mathrm{d}\rho\, \mathrm{d}f\, p(\rho, f|\theta_n)\, p(\bold{x}|\rho,f) \frac{p(\rho,f|\bold{x},\alpha)}{p(\rho,f|\bold{x},\alpha)}\\
&=& p(\bold{x}|\alpha) \times \\
&&\int \mathrm{d}\rho\, \mathrm{d}f\, \frac{p(\rho,f|\theta_n)p(\bold{x}|\rho,f)}{p(\rho,f|\alpha)p(\bold{x}|\rho,f)} p(\rho,f|\bold{x},\alpha)\\
\frac{p(\bold{x}|\theta_n)}{p(\bold{x}|\alpha)}&\approx&\frac{1}{K} \sum^{K}_{k=1} \frac{p(\rho_k, f_k |\theta_n)}{p(\rho_k, f_k |\alpha)}
\end{eqnarray}
which allows us to re-sample our data under the new priors. For a straight comparison between $\theta_1$ and $\theta_2$, we can calculate
\begin{equation}
\frac{p(\bold{x}|\theta_1)}{p(\bold{x}|\theta_2)} \approx \frac{\sum^{K}_{k=1}p(\rho_{k1}, f_{k1} |\theta_1)}{\sum^{K}_{k=1}p(\rho_{k2}, f_{k2} |\theta_2)}.
\label{eq:imp}
\end{equation}

In Figure~\ref{fig:imp_sampling} we show the sampling in dilution and mean stellar density under uniform priors in gray and the two different models, $\theta_1$ and $\theta_2$ in blue and red, respectively. The blue model prior distribution overlaps significantly more with the sampled distribution than the red model, hinting that we are going to strongly prefer this model. We used Equation~\ref{eq:imp} under our modeling and found ${p(\bold{x}|\theta_1)}/{p(\bold{x}|\theta_2)} = 4713$, that is we are confident that the planets orbiting the primary with a confidence level of 99.98\%, provided all planets orbit the same star star.




The reason why we are able to infer the planet's the host star is that the various transiting planet parameters are not entirely degenerate with dilution and stellar density. The key parameters that control the shape of the various transits are the stellar density which decreases with increasing transit duration, the planet-to-star radius ratio which increases with transit depth, and the impact parameter which causes transits to be more 'V'-shaped and decreases transit depths. So, for example, the transit can only be made so short by increasing the impact parameter before it becomes too V-shaped to fit the data: this limits the values stellar density can take. 

\section{Revised parameters with the planets orbiting the primary}
\label{sec:revision}

Given we are very confident that all the planets orbit the primary star (provided they all orbit the same star, this is discussed in detail in Section~\ref{sec:validity}), we can revise the stellar and planetary properties we calculated from our light curve modeling, properly accounting for the effect of dilution from the stellar companion. Given that a posterior probability is proportional to the product of the prior probability and the likelihood we can calculate the system's parameters by weighting the original samples by the marginalized likelihood under the model of the planet orbiting Kepler-296 A where weights are equal to the probability $p(\bold{x}|\theta)/p(\bold{x}|\alpha)$. In Table~\ref{tab:weighted_params} the weighted median and weighted quartiles are reported where a weighted median has 50\% of the weight on either side. In addition to the sampled parameters we report the ratio of the semi-major axis to the stellar radius ($a/R_s$), the semi-major axis ($a$), the planetary radius ($R_p$) and the stellar flux incident on the planets ($S_p$). $a/R_s$ depends only on stellar density and orbital period, while $a$ and $R_p$ rely on the stellar radius. We draw $n$ stellar radii samples from the MCMC stellar property chains to properly include the stellar radius uncertainty and multiply these by  $a/R_s$ and $R_p/R_s$ to infer $a$ and $R_p$, where $n$ is the number of transit model samples. $S_p$ in solar-earth units can be parametrized as a function of $a/R_s$ and $T_s$ so that 
\begin{equation}
S_p/S_\oplus = \left(\frac{R_s}{a} \times \frac{a_\oplus}{R_\odot}\right)^2 \left(\frac{T_s}{T\odot}\right)^4
\end{equation}
where $S_\oplus$ is the incident flux on the Earth from the Sun, $(a_\oplus/R_\odot)$ is the semi-major axis of the Earth in units of stellar radius where we use 215.1 and $T_\odot$ is the effective temperature of the Sun which we have taken as 5778 K.

The five planets have radii of between 1.5 and 2.1 $R_\oplus$, which places places Kepler-296A into an elite group of planets with five small transiting planets that also includes Kepler-62 \citep{borucki13}, Kepler-186 \citep{quintana14} and Kepler-444 \citep{campante15}. Kepler-296A is also one just a handful stars to host a sub-2 $R_\oplus$ planet that receives less incident flux than Earth receives from the Sun (Kepler-296 Af). Kepler-296 Ae and Af have been previously reported as potential habitable zone planets \citep{rowe14,torres15} -- the region around a star where liquid water could exist given favorable atmospheric conditions. \citep{kopparapu13} published theoretical limiting incident flux to fall within a circumstellar habitable zone, we find that Kepler-296 Ae falls into the Kopparapu et al's `optimistic' habitable zone and Kepler-296 Af in their `conservative' habitable zone.

\citet{torres15} model and report parameters for the outer two planets, the habitable zone planets, in the Kepler-296 system. However, Torres et al. provide planet parameters for both stellar host scenarios. The planet parameters they report from their multi-modal nested sampling using \textsc{mulinest} \citep{feroz08,feroz09} are largely consistent with our results if we consider only their model with planets orbiting the larger star. We see somewhat significant differences in Torres et al's derived planetary parameters. The discrepancy can be explained by the different stellar properties Torres et al. assume for the two stars compared with the analysis we present here. Torres et al. use spectroscopic parameters reported by \citet{muirhead14} which used just a single stellar spectrum model. By using two model spectra modeling we are able to improve upon previous stellar properties for both the star's in this system, removing the bias inherent when ignoring the companion.

\begin{figure}[htbp]
\begin{center}
\includegraphics[width=0.48\textwidth]{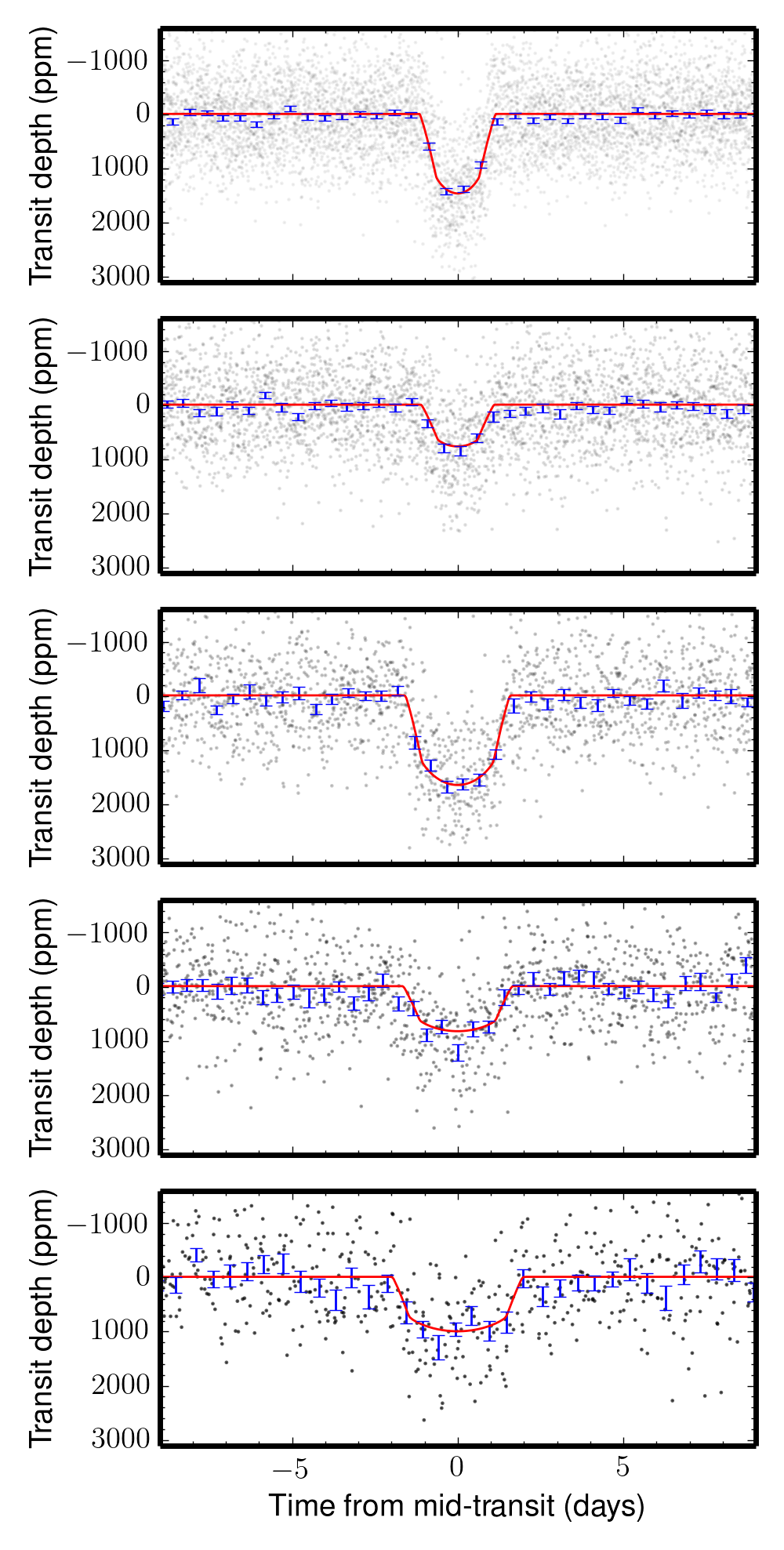}
\caption{Transits of the five planets in the Kepler-296 system. The planets are in order of increasing orbital period from planet b to f. The data have been folded on the best fitting orbital period. The observed data is shown in black and binned data in blue. The best fitting model is shown in red. Note that while we show the binned data, no calculations are performed on these data.}
\label{fig:transitplot}
\end{center}
\end{figure}

\begin{table*}[htdp]
\caption{Inferred stellar and planetary parameters from our MCMC modeling. Parameters are the weighted quartiles of the posterior distribution where the weights were calculated via importance sampling.}
\begin{center}
\begin{tabular}{lllll}
Body&Parameter & W. Med &  84.1\% & 15.9\% \\
\hline
Kepler-296 A	&$\rho$\footnote{The stellar radius and stellar mass for Kepler-296A are listed in Table~\ref{tab:stellar} and are $R_\odot = 0.480\pm0.076$, and $R_\odot = 0.498\pm0.076$. The stellar mass and radius are not strictly consistent with the density present here because this density was calculated using additional information from the transit model.} (g/cc)		&	7.19			&	+0.9	&	-1.0		\\
			&$\gamma_1$		&	0.78			&	+0.20	&	-0.27		\\
			&$\gamma_2$		&	-0.13			&	+0.34	&	-0.26		\\
			&Dilution			&	0.215		&	+0.012	&	-0.012	\\
\hline
Kepler-296 Ac	&Epoch (BKJD\footnote{The time zeropoint we use id the Barycentric Julian Date minus a fixed offset of 2454833 days. This is referred to as BKJD and is the time system used in all Kepler data products.})		&	135.9229		&	+0.0013	&	-0.0013	\\	
			&Period (days)		&	5.8416366	&	+1.0e-5	&	-1.0e-5	\\
			&Impact parameter	&	0.25			&	+0.18	&	-0.16		\\
			&$R_p / R_s$		&	0.0381		&	+0.0014	&	-0.0012	\\
			&$e\cos{\omega}$	&	0.000		&	+0.042	&	-0.052	\\
			&$e\sin{\omega}$	&	-0.000		&	+0.017	&	-0.051	\\
			&Radius ($R_\oplus$)&	2.00			&	+0.33	&	-0.32		\\
			&Incident Flux  ($S_\oplus$)&	14.8		&	+2.7		&	-2.3		\\
			&$a/R_s$			&	23.5			&	+0.9		&	-1.1		\\
			&$a$ (AU)			&	0.0521		&	+0.0088	&	-0.0086	\\
            &eccentricity (3-$\sigma$ upper limit)			& $<$0.33			&		&		\\
\hline
Kepler-296 Ab	&Epoch (BKJD)		&	131.1285		&	+0.0030	&	-0.0039	\\
			&Period (days)		&	10.864384	&	+5.1e-5	&	-4.6e-5	\\
			&Impact parameter	&	0.65			&	+0.09	&	-0.19		\\
			&$R_p / R_s$		&	0.0308		&	+0.0018	&	-0.0021	\\
			&$e\cos{\omega}$	&	0.000		&	+0.086	&	-0.075	\\
			&$e\sin{\omega}$	&	0.01			&	+0.14	&	-0.04		\\
			&Radius ($R_\oplus$)&	1.61			&	+0.29	&	-0.27		\\
			&Incident Flux  ($S_\oplus$)&	6.5		&	+1.2		&	-1.0		\\
			&$a/R_s$			&	35.5			&	+1.4		&	-1.7		\\
			&$a$ (AU)			&	0.079		&	+0.013	&	-0.013	\\
            &eccentricity (3-$\sigma$ upper limit)			&	$<$0.33		&		&		\\
\hline
Kepler-296 Ad	&Epoch (BKJD)		&	133.6496		&	+0.0022	&	-0.0022	\\
			&Period (days)		&	19.850291	&	+6.1e-5	&	-5.7e-5	\\
			&Impact parameter	&	0.26			&	+0.18	&	-0.16		\\
			&$R_p / R_s$		&	0.0398		&	+0.0014	&	-0.0012	\\
			&$e\cos{\omega}$	&	0.000		&	+0.046	&	-0.0042	\\
			&$e\sin{\omega}$	&	-0.001		&	+0.017	&	-0.050	\\
			&Radius ($R_\oplus$)&	2.09			&	+0.33	&	-0.32		\\
			&Incident Flux  ($S_\oplus$)&	2.90		&	+0.52	&	-0.44		\\
			&$a/R_s$			&	53.1			&	+2.1		&	-2.6		\\
			&$a$ (AU)			&	0.118		&	+0.020	&	-0.020	\\
            &eccentricity (3-$\sigma$ upper limit)			&	$<$0.33		&		&		\\
\hline
Kepler-296 Ae	&Epoch (BKJD)		&	136.0350		&	+0.0064	&	-0.0057	\\
			&Period (days)		&	34.14211		&	+0.00025	&	-0.00025	\\
			&Impact parameter	&	0.34			&	+0.19	&	-0.21		\\
			&$R_p / R_s$		&	0.0291		&	+0.0018	&	-0.0015	\\
			&$e\cos{\omega}$	&	0.000		&	+0.053	&	-0.052	\\
			&$e\sin{\omega}$	&	0.000		&	+0.043	&	-0.042	\\
			&Radius ($R_\oplus$)&	1.53			&	+0.27	&	-0.25		\\	
			&Incident Flux  ($S_\oplus$)&	1.41		&	+0.25	&	-0.21		\\
			&$a/R_s$			&	76.2			&	+3.0		&	-3.7		\\
			&$a$ (AU)			&	0.169		&	+0.029	&	-0.028	\\
            &eccentricity (3-$\sigma$ upper limit)			&	$<$0.33		&		&		\\
\hline
Kepler-296 Af	&Epoch (BKJD)		&	162.6069		&	+0.0071	&	-0.0072	\\
			&Period (days)		&	63.33627		&	+0.00060	&	-0.00062	\\
			&Impact parameter	&	0.55			&	+0.11	&	-0.24		\\
			&$R_p / R_s$		&	0.0344		&	+0.0021	&	-0.0019	\\
			&$e\cos{\omega}$	&	-0.000		&	+0.071	&	-0.078	\\
			&$e\sin{\omega}$	&	-0.00			&	+0.10	&	-0.04		\\
			&Radius ($R_\oplus$)&	1.80			&	+0.31	&	-0.30		\\
			&Incident Flux  ($S_\oplus$)&	0.62		&	+0.11	&	-0.10		\\
			&$a/R_s$			&	115.1		&	+4.5		&	-5.6		\\
			&$a$ (AU)			&	0.255		&	+0.043	&	-0.042	\\
            &eccentricity (3-$\sigma$ upper limit)			&	$<$0.33		&		&		\\
\hline
\end{tabular}
\end{center}
\label{tab:weighted_params}
\end{table*}%

\section{The validity of our assumption that the planets orbit the same star}
\label{sec:validity}

An critical assumption made in the analysis so far is that all five planets orbit the same star. However, there is at least one example of a planetary system with planets orbiting different stellar members of the system - Kepler-132 \citep{lissauer14}. In the Kepler-132 system, two planets orbit one stellar companion of a binary while one planet orbits its companion. As a result, it is important that we assess whether our assumption that the planets all orbit the same star is justified.

In order for the planets to orbit different stars, and yet be seen in transit, the orbital planes of the two planetary systems must nearly line up (their angular momentum vectors must point in nearly the same direction). 

In approximate terms, the difference in angle between the two orbital
planes (of the two putative planetary systems), must be less than the
alignment necessary for the planets to transit, i.e., 
\be
\Delta \theta \lta {R_\ast \over a_P} \sim 
{0.3 R_\odot \over 0.2 {\rm AU}} \sim 0.0070 \, rad \sim 
0.4^\circ \,,
\ee
where we have used the small angle approximation. In other words, the
required alignment is extremely tight, with a tolerance of less than a
half a degree. The binary has a projected separation of about 35 AU,
so the question becomes: What is the probability that nature will
produce two planetary systems around the two binary components
separated by at least 35 AU, such that the orbital angular momenta 
point in the same direction (within $0.5^\circ$)?

\subsection{Considerations of Disk Formation} 
\label{sec:form} 

Molecular cloud cores are the sites of star formation and their
angular momentum profiles drive the formation of the circumstellar
disks (which, in turn, provides the sites for planet formation). The
rotation rate of these cores is estimated by measuring the velocity
gradient of a given molecular line across a map of the core (starting
with \citealt{goodman93}). However, the direction of the inferred
(two-dimensional) angular momentum vector varies from point to point
within the map \citep{caselli02}. As a result, the mean velocity
gradients have values of $\sim1-2$ km s$^{-1}$ pc$^{-1}$, but the 
direction of the rotation varies by 10 -- 30 degrees across the
map. Moreover, the emission maps show a coherence length of
$\ell\approx0.01$ pc, i.e., for two points separated by distances
larger than $\ell$, the directions of the rotation vectors are
uncorrelated. Here, ``uncorrelated'' means chosen randomly from 
the distribution of values within the measured range, where the 
measured range is of order 30 degrees. (Note that the range is {\it not} 
360 degrees; if that were the case, the rotation vectors would 
take on a purely random direction for points separated by distances 
greater than $\ell$.) 

We can build a simple model of star/binary/disk formation using 
the results given above. As a first approximation, consider the 
density profile of the initial molecular cloud core to have the 
isothermal form 
\be
\rho = \Lambda {a^2 \over 2 \pi G r^2} \,,
\ee
where $a$ is the isothermal sound speed and $\Lambda\sim1-2$ is the
overdensity factor \citep{fatuzzo04} that accounts for the
observed condensation velocities \citep{lee99}. The corresponding
enclosed mass is thus given by 
\be
M(r) = \Lambda {2 a^2 \over G} r \,. 
\ee
The radius $r_P$ that initial encloses the mass $M_P$ of the primary
can be written in the form
\be
r_P = {G M_P \over 2 \Lambda a^2} \approx 3 \times 10^{16} {\rm cm} 
\left( {M_P \over 0.33 M_\odot} \right) 
\left( {a \over 0.20 {\rm km/s}} \right)^{-2} \,.
\ee
We thus note that $r_p \sim \ell \sim 0.01$ pc, i.e., the sphere that
initially contained the mass of the primary is comparable to the
coherence length observed in molecular cloud cores. As a result, the
primary, and the inner disk that forms its planetary system, can have
a different direction for its angular momentum vector than the
material that collapses later to form the secondary.  Further, we
would expect that the angle between the angular momenta of these
different layers of the core to be 10s of degrees. With this type of
initial conditions, the direction of the orbit of the binary companion
is predicted to differ from that of the planetary system by 10s of
degrees \citep[also see][]{spalding14}. 

As a consistency check, consider the centrifugal radius produced 
during the collapse. When the inner portion of the core has 
collapsed to form the primary, the centrifugal radius 
\be
R_C = {G^3 M_P^3 \Omega^2 \over 16 \Lambda^3 a^8} \approx 
2 - 20 {\rm AU} \, . 
\ee
As a result, the primary and its planetary system, whose size is 
determined by $R_C$ to leading order, can fit inside the observed 
binary orbital separation of the Kepler-296 system (which has 
projected separation of 35 AU and hence expected separation of 
about 70 AU). 

This simple theoretical argument predicts that the disks observed in
binary systems should not be perfectly aligned with the angular
momentum vectors of the binary orbits and that the disks surrounding
the two stars should not aligned with each other.  Since disks
polarize the light scattered from their central stars, polarization
measurements can be used to estimate the angular orientation of disks
on the plane of the sky. Such a study has been carried out for a
collection of 19 binary and higher-order multiple T Tauri systems
\citep{jensen04}; the results show that disks in binary systems are
aligned with each other to within about 20$^\circ$, but are not
exactly coplanar. A similar study for southern star formation regions
\citep{monin06} finds similar results; for 15 binary systems, the
observed angle differences shows a distribution of values, with all
but one in the range 0 -- 40$^\circ$, and more than half of the 
sources showing relatively small angles $\Delta\theta<10^\circ$. 

Both the theoretical argument and the observational studies indicate
that two planetary systems associated with the two members of a binary
pair should be roughly -- but not exactly -- aligned. The range of
possible relative orientation angles appears to be about $\pm 20^\circ$. 
We can thus make a simple estimate of the probability to find highly 
aligned planetary systems: If the relative inclination angle is drawn 
uniformly from the range $-20^\circ < \Delta \theta < 20^\circ$, and
if we need $|\Delta \theta| < 0.5^\circ$ to observe transits, then the
required alignment would occur only about 1 out of 40 times (2.5\%).

\subsection{Probability} 
\label{sec:chain} 

This subsection considers a simple probability argument: The five periods
of the planets are observed to be almost equal spaced in a logarithmic
sense: The period ratios between successive pairs of planets are all
$\sim1.8$ (more precisely, 1.88, 1.83, 1.71, and 1.86, with a mean of
1.82 $\pm$ 0.066). This chain of nearly equal period ratios can
naturally be produced if the planets experienced convergent migration
during their early stages of evolution. On the other hand, if one or
more planets orbit the secondary (with the rest of the planets
orbiting the primary), then this set of nearly equal period ratios
would be highly unlikely.

We can understand (roughly) how convergent migration results in
regular orbital spacing as follows: When multiple planets migrate
within a disk, they often lock into mean motion resonance (MMR) and
move inward together (this phenomenon has been studied by many
authors, starting with \citealt{goldreich65}). Indeed, since orbital
eccentricity is easily excited during migration, the planets must
often be in, or near, MMR to avoid orbit crossing and instability. The
most common and strongest resonances in this context are the 2:1 and
3:2 MMR \citep{murray99}. With these period ratios, the semimajor axes of
the planets are separated by factors $f \approx$ 1.59 and 1.31,
respectively, so that orbital spacings in this range are naturally
produced. Moreover, detailed numerical studies \citep{rein12} indicate
that the period ratios for the multiplanet systems discovered by 
{\it Kepler} are indeed consistent with convergent planet migration. Additional stochastic forces (e.g., due to turbulent fluctuations
in the disk) act to spread out the orbital spacing \citep{adams08}, so that values
in the range 1.3 to 2 are naturally produced (including the value of 
1.8 observed for Kepler-296).

If the five planets detected in association with Kepler-296 do not 
orbit the same star, then there should be no correlation (or anti-correlation)
between the orbital periods of the planets orbiting the two different stars.
Since there are many possible ways for this scenario to be realized, we
illustrate this point here by assuming that one planet orbits the secondary,
while the other four planets orbit the primary. The period of the secondary
planet should be independent of the periods of the four primary planets, so
there is some chance that the secondary planetary period would be close to one
of the periods of the others -- close enough to to render the system apparently 
unstable if one mistakenly considered all of the planets to orbit a single star.

To make a numerical estimate, suppose that the orbital periods are
distributed with a log-random distribution. The observed planets have
periods with a spacing of $\sim1.8$ (as described above). However, if
a planetary pair were to have a period ratio that is too close to
unity, it would most likely be unstable (in the absence of a well-tuned 
resonant state). As a general rule, orbitaly instability sets in when the
semi-major axes are too close, more specifically when their 
separation is less than several mutual Hill radii. In practice, however, 
we find that the Kepler multi-planet systems have extremely few period 
ratios less than 4:3, i.e., the period ratio is almost never
observed to be less than 4/3=1.33.

Suppose we have a chain of planets with a factor of 1.8 spacing in
period and we choose another planet from a log-random distribution.
Then the chances of the new planet being too close to another planet
in the chain is given approximately by the expression 
\be
P = {\ln(4/3) \over \ln(1.8)} \approx 0.49 \,. 
\ee
In other words, about 49 percent of the time, if you choose a planet
around the secondary, it would have a period that is too close to one
of the other planets (orbiting the primary), such that it would
apparently lead to an unstable system.

Note the one could derive a wide range of values for the above 
probability, where the result depends on how you 
define the system and what information you take as given.  Let's now
consider a more extreme case. Suppose that you have a planetary system
with 4 planets orbiting the primary, with the observed factor of
$\sim1.8$ spacing in orbital periods. And then suppose that you choose
a 5th planet to orbit the secondary. In principle the 5th planet can
have any orbital period and be stable. For the sake of definiteness,
we take the allowed range of periods to be the range observed in the
system, i.e., a factor of $\sim11$ in period. If we then require that
the fifth planet continue the chain of period ratios, it must have a
value in the range $1.82 \pm 0.066$ times the period of the fourth
planet. Thus, the probability of the 5th planet continuing the chain
of period ratios is approximately given by
\be
P \approx { \ln(1.886/1.754) \over \ln(11)} \approx 0.030 \, . 
\ee
In other words, the probability of continuing the observed chain 
of orbital period ratios is about 3 percent (and this value would 
be smaller if we allowed for a wider range of orbits to choose from).  

\subsection{Discussion of planetary alignment} 
\label{sec:discuss} 

The two arguments given above suggest that two highly aligned
planetary systems orbiting the two members of a binary pair will be
rare. More specifically, the chances of forming such a system (Section
\ref{sec:form}) and the chances that planets orbiting two stars
produce a coherent chain of period ratios (Section \ref{sec:chain})
are both approximately $2-3\%$. 

Either argument, by itself, is highly suggestive but not definitive.
With a 3\% occurrence rate, if we observe $\sim35$ binary systems
containing multiple transiting planets, then we would expect to find
sufficient alignment in (of order) one system. In this case, however,
the two arguments are independent. We need to see the five planets of
the Kepler-296 system in transit (a 2.5\% effect) and also see the
observed chain of period ratios (another 3\% effect). The chances of
both properties occurring is thus much lower, with a probability
$P\sim(0.025)(0.03)$ = $7.5\times10^{-4}$. With this probability, we
would need to observe more than 1300 binary systems with multiple
planets in order to expect one with these properties.

In the most recent release of Kepler planet candidates, there are 608 multiple transiting planet systems comprising 1492 planet candidates \citep{mullally15}. Of those systems, only Kepler-132 presents a compelling case for a binary star with transiting planets orbiting both stars. The planets in the Kepler-132 system are on considerably shorter periods (6.1, 6.4 and 18 days) and orbit a star more than twice the size than the planets in the Kepler-296 system, hence the probability of the observed planets to transit Kepler-132 is much higher. The probability to transit for Kepler-132c and d is roughly 8\% and 4\%, respectively, whereas the probability for Kepler-296Ae and f to transit is roughly 1.3\% and 0.9\%, respectively. Therefore, it is many times more probable that two stars with have disks aligned enough for transiting planets in the Kepler-132 system than in the Kepler-296 system. Indeed, we shouldn't be surprised to see a false positive case occurring for close-in two-planet systems. 

Perhaps a more fair family to compare Kepler-296 to is the systems with at least 5 transiting planets. Of the 608 multi-transiting planet systems just 24 host at least 5 planet candidates, inclusive of Kepler-296. At most 50\% of these five planet systems are likely to be binaries \citep{wang14}. Taking the false positive rate for systems like Kepler-296 as $7.5\times10^{-4}$ and having (0.5 $\times$ 24) systems, we would expect to see 0.009 false multi-systems, making it very likely that the planets in the Kepler-296 system orbit the same star. This result validates the assumptions made in earlier sections and leads us to the conclusion that the five planets orbit Kepler-296A.

\section{Nomenclature}
\label{sec:nomen}

We want to briefly touch upon our naming system for the planets in this system. The five planets have KOI numbers 1422.01, 1422.03, KOI-1422.02, KOI-1422.05 and KOI-1422.04 in order of increasing orbital period. Owing to the additional data available to us in this work compared to that used by \citet{rowe14} and \citet{lissauer14}, we realized that KOI-1422.03 had an orbital period of 10.86 days, a factor of precisely three longer than was reported by \citet{rowe14} and \citet{lissauer14} who assigned the Kepler number 296 to this system. To retain consistency with \citet{rowe14} we stick with the letter \emph{b} to identify KOI-1422.03 even though planet \emph{b} now has a longer orbital period than planet \emph{c}. We note that recent Kepler planet candidate catalogs all list KOI-1422.03 with an orbital period of 10.86 d \citep{burke14,rowe15,mullally15}. In addition, \citet{torres15} used 10.86 d as the orbital period of KOI-1422.03 in their analysis of the stellar density (G. Torres \& D. Kipping, priv. comm.). All papers subsequent to the earlier discovery papers find an orbital period of 10.86 d and we believe that it is highly probably that this is the correct orbital period solution.

In our naming scheme we also the letter \emph{A} after the primary star name to make it clear that the planets orbit the primary star in the binary. Consequently, identifiers now used for these planets, in order of increasing orbital period are Kepler-296 Ac, Kepler-296 Ab, Kepler-296 Ad, Kepler-296 Ae and Kepler-296 Af. We summarize this information in Table~\ref{tab:nomen}.

\begin{table}[htdp]
\caption{Nomenclature for the Kepler-296 A planets.}
\begin{center}
\begin{tabular}{llll}
\hline
P$_\textrm{orb}$ (d)&\citealt{burke14} & \citealt{rowe14} & This work \\
\hline
5.8		&	KOI-1422.01	&	Kepler-296 c	&	Kepler-296 Ac \\
10.9		&	KOI-1422.03	&	Kepler-296 b	&	Kepler-296 Ab \\
19.9		&	KOI-1422.02	&	Kepler-296 d	&	Kepler-296 Ad \\
34.1		&	KOI-1422.05	& 	Kepler-296 e	&	Kepler-296 Ae \\
63.3		&	KOI-1422.04	&	Kepler-296 f	&	Kepler-296 Af \\
\hline
\end{tabular}
\end{center}
\label{tab:nomen}
\end{table}%

\section{Conclusions}

The Galaxy is expected to contain many binary systems with planets orbiting one or both stellar components. In this work, we have explored whether it is likely that one will observe planets (in multi-planet systems) transiting both stellar components. For the solar system architecture of the Kepler-296 system, this 
exploration strongly indicates that all five of the observed planets are likley to orbit the primary. 

The paper presents a statistical analysis which demonstrates that if the planets all orbit the same star, then that host star must be the brighter component of the binary system. For this analysis, we start by modeling the Kepler light curve of Kepler-296 under a prior with no preference for stellar properties (see Section \ref{sec:lightcurve}). We then compare the two competing models of (a) planets orbiting the brighter star or (b) planets orbiting the fainter star, where the comparison uses importance sampling under the different priors (Section \ref{sec:compare}). We find that the planets are significantly more likely to orbit the brighter star. 

The paper also provides a supporting statistical argument for the orbital alignment of high multiplicity transiting planet systems such as Kepler-296. This analysis (Section \ref{sec:validity}) indicates that it is highly unlikely that one would observe planets transiting both stars --- where the odds are less than about 1 in 1000 for this example. The reason for this low probability of observing planets in transit orbiting both stars is twofold: (a) the probability of randomly observing a system with planets orbiting both stars such that the system is apparently stable and the planets are in an apparent (near) resonant chain is only a few percent, and (b) the probability of having the two stars with aligned circumstellar disks is also only a few percent. The probability of realizing \emph{both} of these independent and unlikely events occurring is thus low. 

The combination of the two analyses described above provides a compelling argument that the currently observed planets in the Kepler-296 system must orbit the brighter star. With this finding taken as given, the planetary properties can be determined to greater precision than before. This paper thus reports revised estimates for the planetary radii, orbital periods, and incident stellar fluxes for the outer planets in the system Kepler-296 Ae and Kepler-296 Af (see Section \ref{sec:revision}). This update to the planetary parameters is important, as these planets are thought to orbit within or near the habitable zone (see also \citealt{torres15}) of their host star (argued here to be the primary, Kepler-296A). After the first discovery of the first Earth-sized planet in the habitable zone \citep{quintana14}, and a dozen subsequent detections \citep{torres15}, we expect the Galaxy to be brimming with analogous planets. The discovery and characterization of such planetary systems thus poses a rich problem for future work.

\acknowledgements
This paper includes data collected by the \K{} mission. Funding for the \K{} mission is provided by the NASA Science Mission Directorate. We would like to express out gratitude to all those who have worked on the Kepler pipeline over the many years of the Kepler mission. Some \K{} data presented in this paper were obtained from the Mikulski Archive for Space Telescopes (MAST) at the Space Telescope Science Institute (STScI). STScI is operated by the Association of Universities for Research in Astronomy, Inc., under NASA contract NAS5-26555. Support for MAST for non-HST data is provided by the NASA Office of Space Science via grant NNX09AF08G and by other grants and contracts. This research has made use of the NASA Exoplanet Archive, which is operated by the California Institute of Technology, under contract with the National Aeronautics and Space Administration under the Exoplanet Exploration Program. This research made use of APLpy, an open-source plotting package for Python hosted at \url{http://aplpy.github.com}. This work was supported by a NASA Keck PI Data Award, administered by the NASA Exoplanet Science Institute. Some of the data presented herein were obtained at the W. M. Keck Observatory from telescope time allocated to the National Aeronautics and Space Administration through the agency's scientific partnership with the California Institute of Technology and the University of California. The Observatory was made possible by the generous financial support of the W. M. Keck Foundation. The authors wish to recognize and acknowledge the very significant cultural role and reverence that the summit of Maunakea has always had within the indigenous Hawaiian community. We are most fortunate to have the opportunity to conduct observations from this mountain. E.V.Q. is supported by a NASA Senior Fellowship at the Ames Research Center, administered by Oak Ridge Associated Universities through a contract with NASA. B.T.M. is supported by the National Science Foundation Graduate Research Fellowship under Grant No. DGE‐1144469. D.H. acknowledges support by the Australian Research Council's Discovery Projects funding scheme (project number DE140101364) and support by the National Aeronautics and Space Administration under Grant NNX14AB92G issued through the Kepler Participating Scientist Program.
\bibliographystyle{apj}
\bibliography{biblio,references,rats-new-new}


\end{document}